\documentclass[conference]{IEEEtran}
\usepackage[pdftex]{graphicx}
\usepackage{graphicx}
\usepackage[dvips]{epsfig}

\usepackage[caption=false,font=footnotesize]{subfig}

\usepackage{amsmath}

\usepackage{cite}

\usepackage{url}

\usepackage{siunitx}

\usepackage{lipsum}

\usepackage[mathscr]{euscript}
\usepackage{booktabs}
\usepackage{comment}
\usepackage[flushleft]{threeparttable} 

\usepackage{xcolor,colortbl}
\usepackage{color,soul}

\definecolor{morange}{rgb}{0.8,0.2,0}
\definecolor{mblue}{rgb}{0,0.3,1.0}
\definecolor{mpink}{rgb}{1.0,0.6,0.6}
\definecolor{mgreen}{rgb}{0.1,0.6,0.2}
\definecolor{mgoodgreen}{rgb}{0.9,1.0,0.7}

\definecolor{Gray}{gray}{0.85}

\usepackage{multirow}
\usepackage{array}
\newcolumntype{L}[1]{>{\raggedright\let\newline\\\arraybackslash\hspace{0pt}}m{#1}}
\newcolumntype{C}[1]{>{\centering\let\newline\\\arraybackslash\hspace{0pt}}m{#1}}
\newcolumntype{R}[1]{>{\raggedleft\let\newline\\\arraybackslash\hspace{0pt}}m{#1}}

\newcolumntype{G}{>{\columncolor{mgoodgreen}}c}

\begin{document}

 \title{Demo: A Transparent Antenna System for \\In-Building Networks}

\author{
\IEEEauthorblockN{Sang-Hyun Park, Soo-Min Kim, Seonghoon Kim$^*$, HongIl Yoo$^*$, Byoungnam Kim$^*$, and Chan-Byoung Chae}
\IEEEauthorblockA{School of Integrated Technology, Yonsei University, Korea\\
$^*$Sensor View, Korea\\
Email:\{williampark, sm.kim, cbchae\}@yonsei.ac.kr, \{seonghoon.kim, hi.yoo, klaus.kim\}@sensor-view.com}
}

\vspace{-10pt}
\maketitle
\begin{abstract}
For in-building networks, the potential of transparent antennas, which are used as windows of a building, is presented in this paper. In this scenario, a transparent window antenna communicates with outdoor devices or base stations, and the indoor repeaters act as relay stations of the transparent window antenna for indoor devices. 
At indoor, back lobe waves of the transparent window antenna are defined as interference to in-building networks. Hence, we analyze different SIR and SINR results according to the location of an indoor repeater through 3D ray tracing system-level simulation. Furthermore, a link-level simulation through a full-duplex software-defined radio platform with the fabricated transparent antenna is presented to examine the feasibility of the transparent antenna.
\end{abstract}


\IEEEpeerreviewmaketitle


\section{Introduction}
Due to the increasing number of Internet of Things (IoT) devices and the emerging vehicle-to-everything (V2X) trend, faster and more reliable networks are required in urban environments.
Installing additional access points and signal repeaters embedded into the existing infrastructure by inserting transparent antennas into windows would be an alternative solution to support the requirements~\cite{1}.
Fortunately, transparent antennas have been developed enough to be used as windows instead of glass via either meshed conductors or transparent conductive oxides (TCOs) for two decades~\cite{2}.
Locating transparent antennas on exterior windows of buildings has benefits: Aesthetic-friendly building, numerous antennas located at various heights, a window antenna serving as a road-side unit (RSU) for multi-objective, and in-building networks aided transparent window antennas. 

In Fig. 1, in the case of in-building networks, the indoor repeaters, which relay the signals to/from the transparent antenna which acts as windows, exist and communicate with indoor devices and users. These repeaters for in-building networks can achieve a higher maximum sum rate to resolve the low indoor signal-to-noise ratio (SNR) caused by blockages, doors, and walls. Besides, indoor devices and users attain lower uplink power consumption than directly transmitting to the base station outside the building. 
Although these advantages exist, the transparent antenna's back lobe waves affect in-building networks as interference. Hence, it is necessary to analyze different signal-to-interference ratio (SIR) and signal-to-interference-plus-noise ratio (SINR) results depending on the location of an indoor repeater through 3D ray tracing system-level simulation.

\begin{figure}[t]
	\centerline{\resizebox{1\columnwidth}{!}{\includegraphics{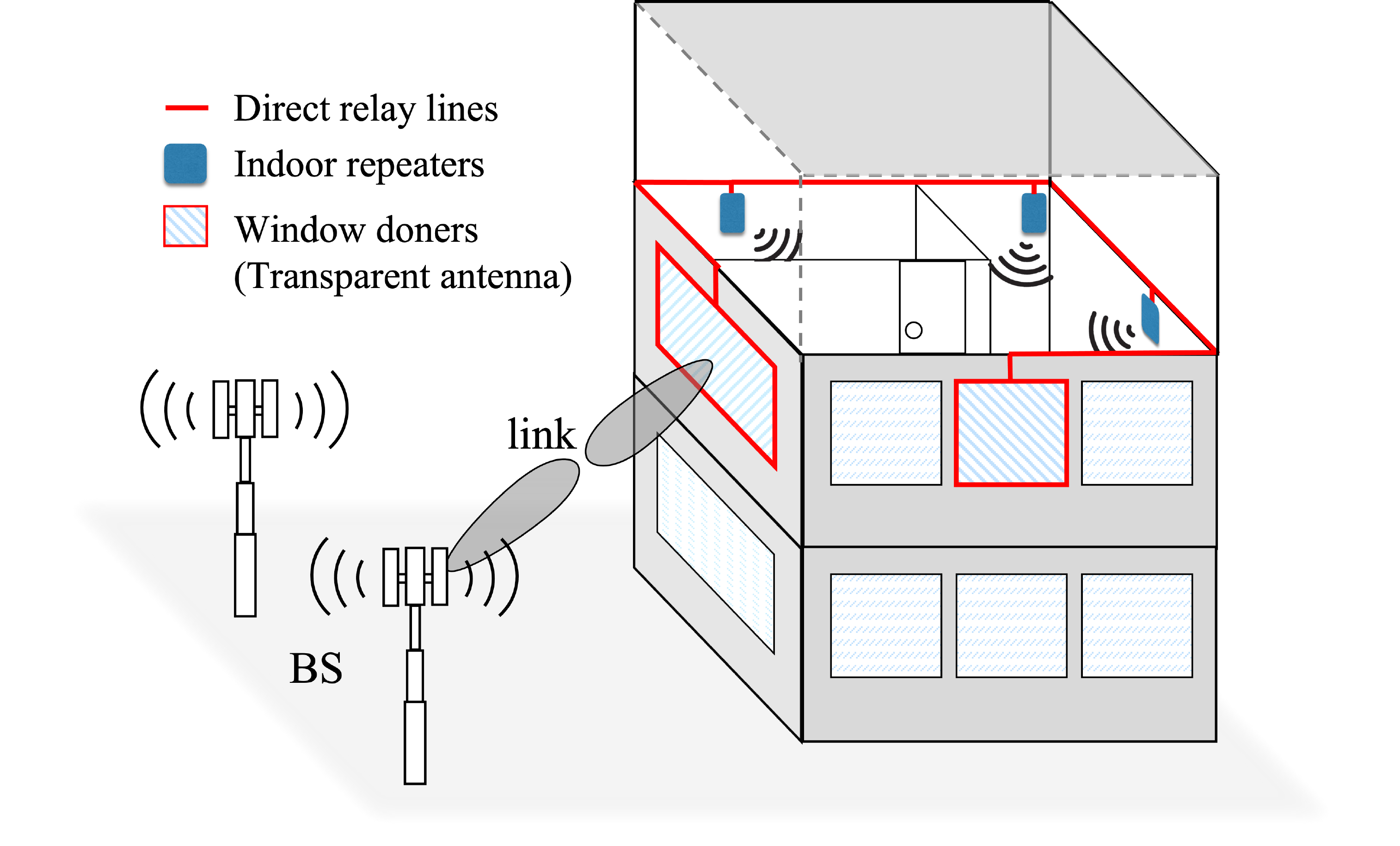}}}
	\caption{An example of transparent antenna system for in-building networks.}
\vspace{-5pt}
\end{figure}

\section{System-level Simulation}

We evaluate SIR and SINR of in-building networks with a system-level simulator using Wireless System Engineering (WiSE)--a 3D ray tracing tool developed by Bell Labs~\cite{KLM+16}.
We model a 3.5 x 3 x 3 m building with a 1.75 x 1.5 m transparent antenna as a window (see Fig. 2(a)). The walls of the building, including a roof, consist of concrete. 
We consider one donor (transparent window antenna) and one repeater in each case. Each case is divided in terms of the different locations of a repeater. The donor's beam always points to the outside of the window. Each indoor repeater's beam points to the center coordinates of the room. 

Fig. 2(b) presents 2D indoor SIR evaluated at 1.5 m height, the middle height of the entire height, in dB scale.
The repeater and the widow's locations are expressed in blue and white squares, respectively. Fig. 2(c) shows the CDF of SINR and SNR graphs based on Fig. 2(b) results in each case. According to results, in case 1, because the locations of the repeater and donor are not a line-of-sight (LoS), in-building networks can support a better maximum sum rate than that from case 2. Hence, when utilizing a window donor for in-building networks, the back lobe wave of the donor that acts as interference and the location of the indoor repeater should be considered necessarily.

\begin{figure*}[t]
	\centerline{\resizebox{2\columnwidth}{!}{\includegraphics{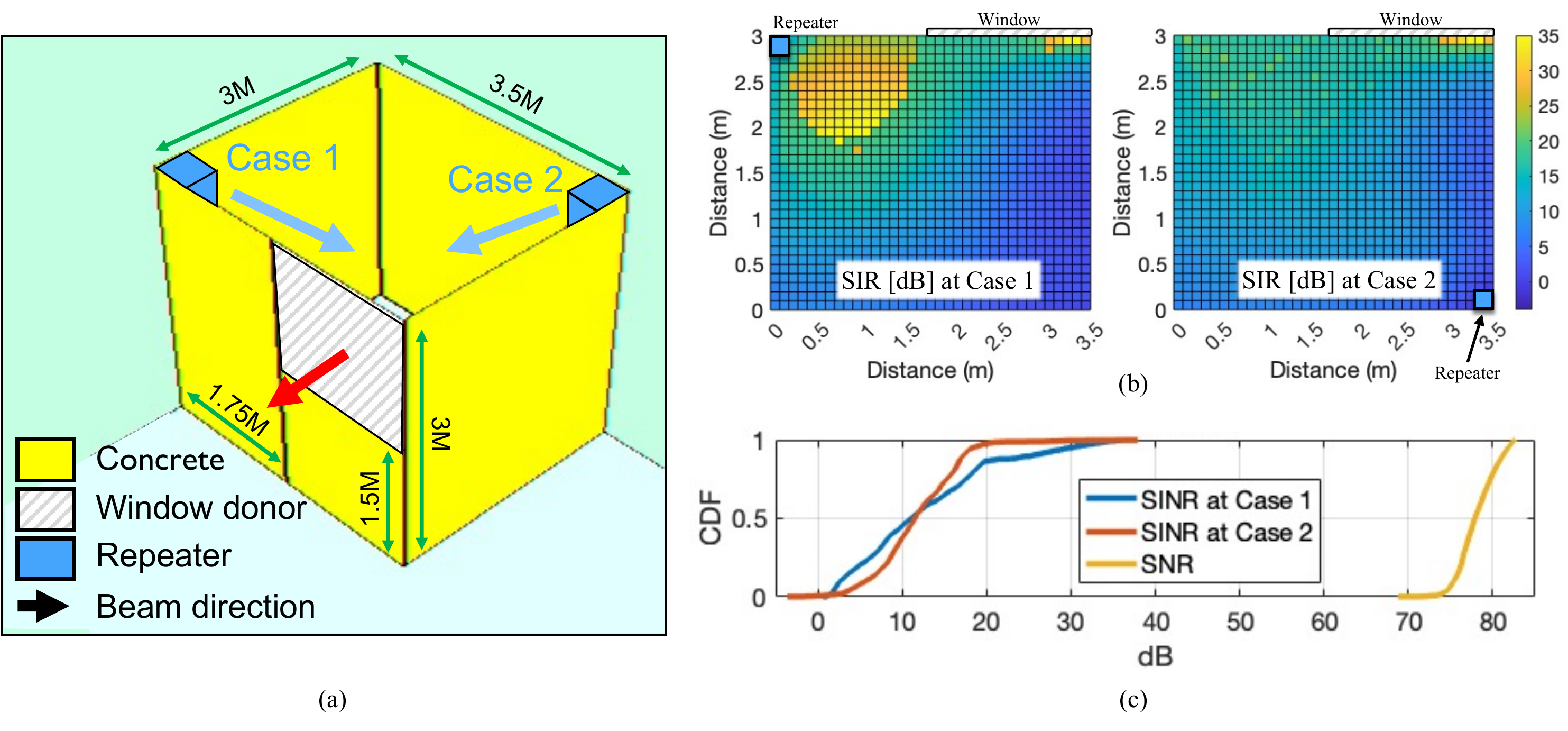}}}
	\vspace{-5pt}
	\caption{(a) In-building networks modeling for 3D ray tracing (b) Indoor SIR  (c) Indoor SINR and SNR graphs.}
	\label{Fig.Rx_panel}
\vspace{-5pt}
\end{figure*}

\section{Demonstration and concluding remarks}
\begin{figure}[]
\centering
\subfloat[\label{Fig.link_fig}]{{\resizebox{0.83\columnwidth}{!}{\includegraphics{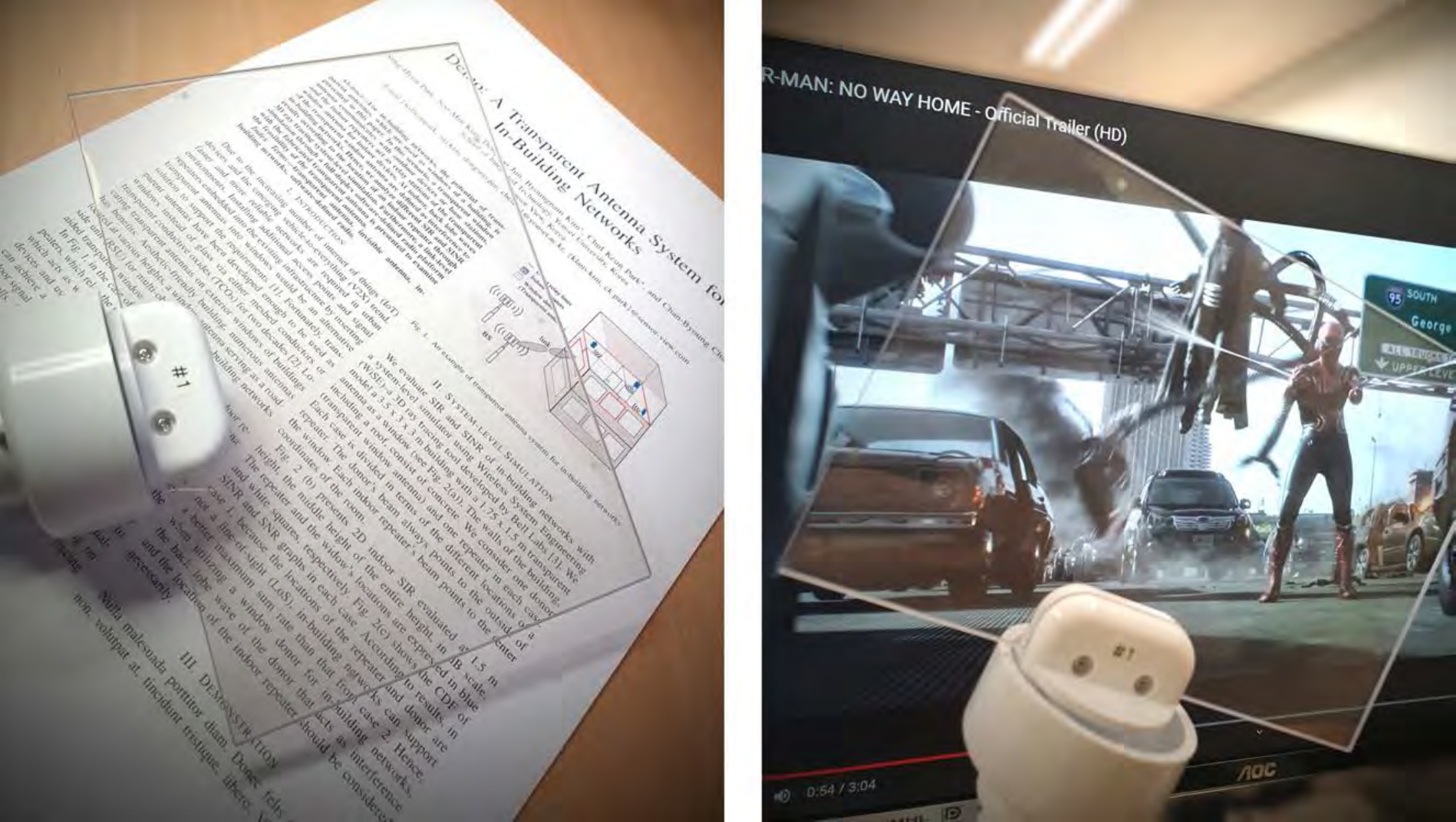}}}}
\hfill
\subfloat[\label{Fig.lens}]{{\resizebox{0.83\columnwidth}{!}{\includegraphics{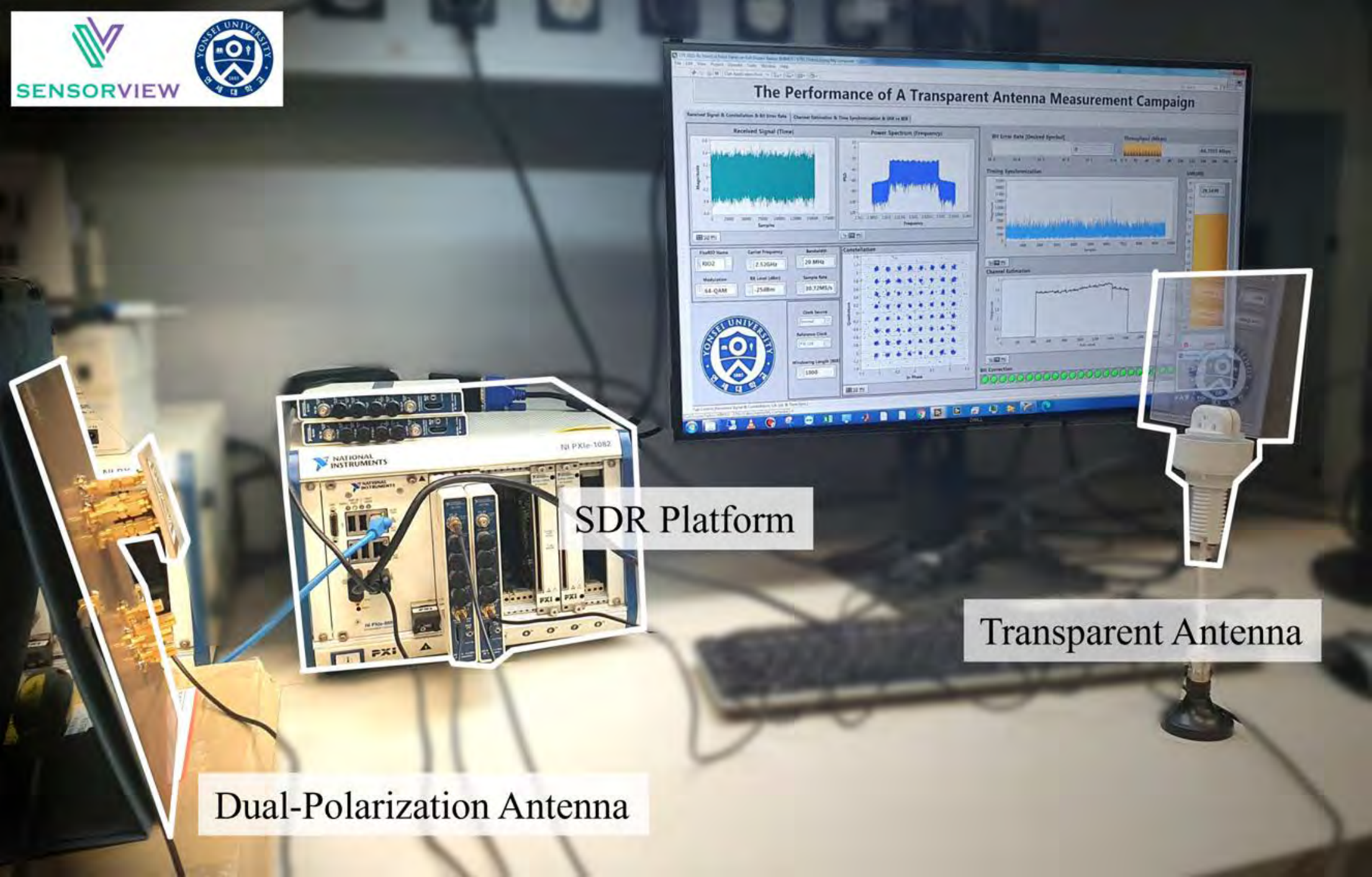}}}}
\hfill
\caption{(a) Proposed transparent antenna (b) real-time measurement campaigns}
\label{fig:dummy}
\vspace{-10pt}
\end{figure}

As Fig.~\ref{fig:dummy} (a) shows, we manufactured a transparent antenna as window donors. Our fabricated sub-6 GHz transparent antenna has high-quality transparency that makes aesthetic-friendly buildings and small back lobe waves that feebly affect the desired signal. To verify and characterize the link-level performance of our proposed antenna with several scenarios, we conducted the measurement campaign as shown in Fig.~\ref{fig:dummy} (b). We tested the scenarios through the FPGA modules, which conclude a PXIe chassis (PXIe-1082), an FPGA controller module (PXIe-8880), RF modules (NI 5791, dual-polarized antenna), and FPGA modules (PXIe-7975). An extended draft will describe more detailed parameters like the antenna's transparency, beam pattern, and hardware specifications.


This paper proposed an in-building networks-aided transparent antenna system. Due to interference from a transparent window antenna, we analyzed indoor SIR and SINR.
As a result, the indoor SINR performance varies critically depending on the location of the indoor repeater. Hence, in order to realize the system, the donor's interference and the indoor repeater's location must be considered.
Furthermore, a link-level simulation through a full-duplex software-defined radio platform with the fabricated transparent antenna was presented to examine the feasibility of a transparent antenna.

\section*{Acknowledgment}
This work was supported by Institute of Information \& communications
Technology Planning \& Evaluation (IITP) grant funded by the Korea government (MSIT) (No. 2021-0-00486)(No. 2021-0-02208).


\bibliographystyle{IEEEtran}

\end{document}